\documentclass[twocolumn,preprintnumbers,amssymb,amsmath,9pt, superscriptaddress]{revtex4}[10pt]
\usepackage{graphicx}
\usepackage{dcolumn}
\usepackage{bm}
\begin{document} 
\input epsf.tex
\newcommand{\beq}{\begin{eqnarray}}
\newcommand{\eeq}{\end{eqnarray}}
\newcommand{\nn}{\nonumber}
\def\ltap{\ \raise.3ex\hbox{$<$\kern-.75em\lower1ex\hbox{$\sim$}}\ }
\def\gtap{\ \raise.3ex\hbox{$>$\kern-.75em\lower1ex\hbox{$\sim$}}\ }
\def\CO{{\cal O}}
\def\CL{{\cal L}}
\def\CM{{\cal M}}
\def\tr{{\rm\ Tr}}
\def\CO{{\cal O}}
\def\CL{{\cal L}}
\def\CM{{\cal M}}
\def\mpl{M_{\rm Pl}}
\newcommand{\bel}[1]{\be\label{#1}}
\def\al{\alpha}
\def\bt{\beta}
\def\eps{\epsilon}
\def\eg{{\it e.g.}}
\def\ie{{\it i.e.}}
\def\mn{{\mu\nu}}
\newcommand{\rep}[1]{{\bf #1}}
\def\be{\begin{equation}}
\def\ee{\end{equation}}
\def\bea{\begin{eqnarray}}
\def\eea{\end{eqnarray}}
\newcommand{\eref}[1]{(\ref{#1})}
\newcommand{\Eref}[1]{Eq.~(\ref{#1})}
\newcommand{\gsim}{ \mathop{}_{\textstyle \sim}^{\textstyle >} }
\newcommand{\lsim}{ \mathop{}_{\textstyle \sim}^{\textstyle <} }
\newcommand{\vev}[1]{ \left\langle {#1} \right\rangle }
\newcommand{\bra}[1]{ \langle {#1} | }
\newcommand{\ket}[1]{ | {#1} \rangle }
\newcommand{\ev}{{\rm eV}}
\newcommand{\kev}{{\rm keV}}
\newcommand{\Mev}{{\rm MeV}}
\newcommand{\gev}{{\rm GeV}}
\newcommand{\tev}{{\rm TeV}}
\newcommand{\mev}{{\rm MeV}}
\newcommand{\meV}{{\rm meV}}
\newcommand{\mnu}{\ensuremath{m_\nu}}
\newcommand{\nnu}{\ensuremath{n_\nu}}
\newcommand{\mlr}{\ensuremath{m_{lr}}}
\newcommand{\acc}{\ensuremath{{\cal A}}}
\newcommand{\mav}{MaVaNs}
\newcommand{\disc}[1]{{\bf #1}} 
\newcommand{\h}{{\cal H}}
\newcommand{\hb}{{\cal \bar H}}

\title{New Matter Effects and BBN Constraints for Mass Varying Neutrinos}
\author{Neal Weiner}
\affiliation{Center for Cosmology and Particle Physics,
  Dept. of Physics, New York University,
New York, NY 10003}
\author{Kathryn Zurek}
\affiliation{Institute for Nuclear Theory and Dept. of Physics,
  University of Washington,
Seattle, WA 98195}

\preprint{INT-PUB 05-21}
\date{\today}
\begin{abstract}

The presence of light $(m_\acc \sim 10^{-6} \ev)$ scalar fields in the early universe can modify the cosmology of neutrinos considerably by allowing their masses to vary on cosmological times. In this paper, we consider the effect of Planck-suppressed couplings of this scalar to electrons and show that such couplings can easily make new sterile states thermally inaccessible in the early universe, preserving the successes of big bang nucleosynthesis predictions. We consider the circumstances under which these effects give the proper initial conditions for recently considered models of neutrino dark energy, and consider limits from tests of the equivalence principle. The parameters which satisfy cosmological constraints naturally give rise to interesting signals in terrestrial neutrino oscillation experiments.

\end{abstract}

\maketitle

\section{Introduction}

One of the most exciting discoveries in recent years has been the demonstration of the existence of neutrino oscillations. The angular flavor dependence of atmospheric $\mu$-neutrinos \cite{Fukuda:1998mi} gave solid evidence to the phenomenon, which was followed up by controlled terrestrial experiments, such as K2K \cite{super} and KamLAND \cite{Eguchi:2002dm}. This, together with the SNO results \cite{Ahmed:2003kj}, has now solidly established neutrino oscillations as the explanation of the long standing solar neutrino problem\cite{Cleveland:1998nv,Fukuda:1996sz, Hampel:1998xg,Altmann:2000ft,Abdurashitov:1999zd}.

However, many questions remain. Namely, what is the origin of neutrino mass and what sets the scale? Are they Dirac or are they Majorana in nature? What is the impact of neutrino mass on cosmology?

One exciting possibility is that neutrino mass is associated with a new, light, scalar field. This is quite reasonable, given that other fundamental fermions have mass arising from the Higgs boson, and that radiative corrections to the mass of such a scalar can be controlled \cite{Fardon:2003eh}. New scalar forces for neutrinos have been considered for many years \cite{Wolfenstein:1977ue,Stephenson:1996qj}. It has been shown that such a field could be the source of all neutrino mass within matter \cite{Sawyer:1998ac}. Additionally, others have considered the possibility that a slowly rolling scalar field with mass $10^{-33}\ev$ might change the mass of neutrinos on Hubble times \cite{Hung:2000yg}, possibly impacting questions of leptogenesis \cite{Gu:2003er}.

Recently, it has been proposed that the dark energy might arise due to interactions between relic neutrinos \cite{Fardon:2003eh}. Here, no Hubble mass particles need be invoked, but rather particles with masses in the range $10^{-6}\ \ev - 10^{-8}\ \ev$. More recently, supersymmetric models have been presented as a means to provide theories where all masses and couplings take on natural values \cite{Fardon:2005wc}. (See also \cite{Takahashi:2005kw} for an alternative supersymmetric model.) One exciting possibility is that these new forces might allow us to test dark energy in neutrino oscillation experiments arising from medium dependent mass values \cite{Kaplan:2004dq}. Studies show that there is still a great deal of room within neutrino experiments for such effects \cite{Zurek:2004vd}, with some studies arguing for possibly interesting effects on solar neutrinos \cite{Barger:2005mn,Cirelli:2005sg}.

\subsection{The early universe problem of sterile neutrinos}

Models of this sort almost always involve the presence of a singlet neutrino, which, at present times, for naturalness reasons, has a mass of $O(\ev)$. The introduction of such a particle forces us to consider the early universe behavior of the theory, as such a neutrino is generally populated in the early universe, leading to severe constraints from big bang nucleosynthesis \cite{Dolgov:2003sg}.

Moreover, the neutrino models of dark energy require that the relic neutrinos be mass eigenstates, not interaction eigenstates, which raises another question of initial conditions.

In this letter, we shall see that the the presence of a new neutrino scalar force, coupled with near-gravitational strength to matter, can allow a resolution of the BBN constraints on sterile neutrinos by changing the mass parameters in the early universe so as to prevent oscillations. We shall also see that the evolution of this setup naturally yields the late time universe as populated with mass eigenstate neutrinos, providing suitable initial conditions for neutrino dark energy.

\section{Neutrino scalar forces}

The simplest means of generating a new force for neutrinos is by the inclusion of a new, singlet neutrino and an associated Yukawa force.
\be
{\cal L}\supset m_D \nu n + \lambda \acc n n + V_0(\acc),
\ee
where $\nu$ is the standard model left-handed neutrino, $n$ is a sterile neutrino, and $\acc$ is the scalar acceleron field.

If the relic neutrinos are in the light mass eigenstate, then there is a contribution to the effective potential of the acceleron at late times
\be
\delta V_{eff} = n_\nu \frac{m_D^2}{\lambda \acc},
\ee
for non-relativistic neutrinos and
\be
\delta V_{eff} = \frac{T^2 m_D^4}{24 \lambda^2 \acc^2}
\label{relmassEigen}
\ee
for a thermal background, both of which will drive \acc\ to large values, in order to minimize the energy of the neutrino background.

However, in the early universe, thermal scatterings produce interaction eigenstates, not mass eigenstates, as long as the sterile state is light enough to be produced (i.e., $\lambda \acc = m_n < T$). We must ask the question: assuming the background neutrinos are presently in mass eigenstates, at what temperature would the sterile state become thermally accessible? This was studied in \cite{Fardon:2003eh} in the case that $V_0(\acc) = \mu^4 \log(\acc/\acc_0)$ and there it was found that $\lambda \acc(T)>T$. However, we would like to explore theories with quadratic potentials, both because quadratic potentials are very generic, and because this is the form of the potential in the recently proposed hybrid models of neutrino dark energy.


Thus, we take a potential $V_0(\acc) = m_\acc^2 \acc^2/2$, and find the background value of the acceleron field
\be
\lambda \acc= \sqrt{\frac{\lambda T}{m_\acc}}\frac{m_D}{12^{1/4}}.
\ee

For $\lambda \acc >T$, we find $\lambda m_D^2/\sqrt{12} m_\acc >T$. Since $m_\acc \sim \lambda m_D$ by naturalness arguments \cite{Fardon:2005wc}, the effects of relativistic neutrinos are generally insufficient to keep the sterile neutrino out of the effective theory at temperatures relevant for BBN.

In order to have a background of mass eigenstate neutrinos,  and to prevent the thermalization of the sterile states, we must have either decohering scatterings  out of equilibrium, such as at $T<\mev$, or $\lambda \acc>T$, so that the heavy state cannot be produced by such scatterings. As we have found that the latter condition cannot be maintained by neutrinos alone, we are prompted to consider other contributions to the effective potential in the early universe.

\subsection{Matter effects and the early universe}

Neutrinos are unique, in that they can feel the effects of  \acc\
through mixing to light states. In contrast, no other fermion in the
standard model can mix with a gauge singlet, even after electroweak
symmetry breaking. However, higher dimensional couplings between \acc\ and these other fermions are not only allowed, they are naturally to be expected from Planck-scale physics. At temperatures near $T\sim 1\ \mev$, only electrons are still in equilibrium, and so we consider  couplings of the form

\be
{\cal L}\supset m_e (1- \frac{\beta_e \acc}{\mpl}) e_l e_r.
\ee

Such a term would arise from the higher dimension operator $\beta_e y_e \acc h e_l e_r/\mpl$ where $h$ is the standard model Higgs field. This is precisely the sort of operator considered in \cite{Sawyer:1998ac,Kaplan:2004dq,Zurek:2004vd}, which lead to new matter effects in neutrino oscillations.

One might think that interactions at this strength (comparable to gravitational), may not have significant effects on the phenomenology of neutrino oscillations in the early universe.  However, due to the enhancement of the high temperature, there can easily be an important correction to the free energy of the acceleron field,
\be
\delta F(\acc) = g_e \frac{T^4}{2\pi^2} I_+ \left(\frac{m_e(\acc)}{T}\right)
\ee
where
\be
I_+(z)=-\int_0^\infty \ dy y^2 \log(1+e^{-\sqrt{y^2+z^2}})
\ee
and $g_e=4$ is the number of degrees of freedom in a Dirac electron. At high temperatures ($T\gg m_e$), this simplifies considerably, yielding a total potential
\be
 V = \frac{T^2 m_e^2}{12} \left( 1 - \beta_e \frac{\acc}{\mpl} \right)^2+\frac{T^2 m_\nu^2(\acc)}{24}+ \frac{m_\acc^2}{2}\acc^2.
\ee

The minimum of the thermal effective potential is at
\be
\acc = \frac{m_e^2 T^2 \beta_e}{6 m_\acc^2 \mpl}=3\ \mev \ \beta_e \left(\frac{T}{1\ \mev}\right)^2\left(\frac{10^{-6}\ \ev}{m_\acc}\right)^2 \ee

To satisfy BBN constraints, we need to ensure that only one state is thermalized. As we have previously stated, this is most straightforwardly achieved by the requirement that the heavy state is thermally inaccessible when electron annihilation to neutrinos is still in equilibrium, i.e.,
 $\lambda \acc(T_{ann}) > T_{ann}$. If we additionally want to produce the proper initial conditions for neutrino dark energy, the heavy state must be inaccessible until after all scatterings have ceased, that is, $\lambda \acc(T_{scat}) > T_{scat}$, which results in the requirement that
\be
3 \beta_e \lambda \left(\frac{T}{\mev} \right) \left( \frac{10^{-6} \ev}{m_\acc} \right)^2 \gtrsim 1.
\label{couplinglimits}
\ee
The annihilation process $\nu_a \bar \nu_a \leftrightarrow e^+ e^-$ freezes out at $T_e \approx 3.2 \mev$ and $T_{\mu\tau} \approx 5.3$ \cite{Dolgov:2003sg}. The scattering process $\nu_a e \leftrightarrow \nu_a e $ freezes out at $1.4 \mev$.

However, the coupling to electrons cannot be arbitrarily large. Such a field generates a Yukawa potential between two atoms of the form
\be
V(r) = -\frac{Z_1 Z_2 G_n m_e^2 \beta_e^2}{4 \pi r} e^{-m_\acc r}.
\ee
$\beta_e$ must be small enough to be consistent with tests of the
gravitational inverse square law (ISL) and equivalence principle
violation limits, as the force couples to lepton number L (or, equivalently, Z). These limits are typically quoted in terms of the
parameter $\alpha$, which in our system is \be
\alpha = \frac{\beta_e^2 m_e^2}{4 \pi m_u^2}
\ee
where $m_u$ is the atomic mass unit.

For a force in the range $m_\acc \simeq 10^{-7} \ev$, equivalence
principle tests yield the tightest constraint, $\alpha \lesssim
10^{-6.5}$ or
\be
\beta_e \lesssim 4 \ee
This limit is roughly constant with masses down to about $10^{-12}\ev$. Slightly heavier accelerons ($m_\acc > 10^{-6}\ev$) have weaker limits on $\beta_e$, which are ultimately dominated by ISL experiments at above $\sim 10^{-5}\ev$. Since the \acc\ expectation value in the early universe drops like $m_\acc^2$, these heavier accelerons tend to have significantly smaller expectation values in spite of the larger allowed values of $\beta_e$. Note that we have ignored the possibility of chameleon effects \cite{Khoury:2003aq,Khoury:2003rn,Gubser:2004du}, which could weaken the limits. Regardless, within these constraints, eqn.~\ref{couplinglimits} is easily satisfied for $T \sim \mev$.

Hence we see that even very weak (Planck-scale) interactions with electrons can drive the acceleron to large values, preventing the thermalization of sterile neutrinos in the early universe, and leaving the background neutrinos as mass eigenstates at late times.

\subsection{Naturalness}

The recent SUSY hybrid models have given a set of models where the neutrino dark energy has all mass parameters of their technically natural size \cite{Fardon:2005wc}. It is important to consider whether the scenario described here can be satisfied in the context of a natural model.

In these models, one can calculate the quantum corrections to the acceleron mass explicitly. Because the loops involve light fields, the masses run all the way from the weak scale to the neutrino mass scale, where the natural size of the acceleron mass is  $m_\acc^2\sim \lambda^2 m_D^2$. At these intermediate temperatures, the natural value is $m_\acc^2\sim \lambda^2 m_D^2/3$. Expressing the previous limit with the replacement $m_\acc^2 = \lambda^2 m_D^2/3$, the earlier limit reads

\be
\frac{ 10^{-3} \beta_e}{ \lambda} \left(\frac{T}{\mev} \right) \left( \frac{10^{-4} \ev}{m_D} \right)^2 >1.
\ee

Notice that this expression can be satisfied for any $m_D$ by taking $\lambda$ sufficiently small. $m_D \simeq 10^{-4}\ev$ was shown to be of the appropriate size to explain the current dark energy.

\subsubsection{Multiple Sterile Neutrinos}

One can incorporate the masses of the heavier neutrinos into these theories by the inclusion of additional sterile states with somewhat larger Dirac masses. However, continuing with the naturalness arguments, these neutrinos with larger Dirac masses should have smaller couplings to the acceleron. As a consequence, their masses are smaller for the same value of \acc, and may begin to thermalize earlier.

Considering for instance the model of reference \cite{Fardon:2005wc}, we have $\lambda \lsim 10^{-5.5}$, $m_D\sim 10^{-1.5}\ev$ and $m_\acc \sim 10^{-7}\ev$. The sterile state can come into thermal equilibrium at $T\sim 70 \mev$ for $\beta_e = 4$.

Once the sterile state can be produced, we must use a density matrix formalism to describe the dynamics, where
\be
\rho = \begin{pmatrix}
\rho_{aa} & \rho_{as} \cr \rho_{sa} & \rho_{ss}
\end{pmatrix}
\ee

If $\rho_{ss}$ begins to be populated, it contributes to the effective potential of the acceleron an amount $\rho_{ss}\lambda^2 T^2/24$. Hence, even for small values of $\rho_{ss}$, this can dominate over $m_\acc^2$ in the potential, driving \acc\ to smaller values, ultimately allowing all sterile states to become populated. Therefore, we should ensure that this cannot happen until after $T=5.34\ \mev$, when $\mu,\tau$-neutrino production freezes out.

There are a number of simple possibilities to achieve this. First, if $m_\acc$ is tuned at the few percent level, we could accommodate $\lambda \sim 10^{-4}$ for $m_D\sim 10^{-1.5}$, which addresses the problem. A second possibility is just that both $\lambda$ and $m_\acc$ are smaller than what saturates the model in \cite{Fardon:2005wc}. For instance, one could take $m_D \sim 10^{-2}\ev$, $\lambda \sim 10^{-7}$, $m_\acc \sim 10^{-9}\ \ev$ (evaluated at the neutrino mass scale) and $\beta_e = 1$.

Lastly, we note one additional interesting possibility. Although muons become non-relativistic much earlier than BBN, their couplings to the acceleron are far less constrained, and can have a significant effect even at temperatures as low as $T\sim 5\ \mev$. If the muons have a non-renormalizable coupling $4 \pi h \mu_l \mu_r \acc/M_{*}$, then
\bea
\frac{\partial V}{\partial \acc} = \frac{2 T^4}{\pi^2} \frac{\partial I_+(m_\mu(\acc)/T)}{\partial \acc}+m_\acc^2 \acc \\ \nonumber
= -\frac{10^{-2}\ \mev^4}{M_*} + m_\acc^2 \acc
\eea
which yields \be
\acc(T=5.34\ \mev) = 10^6\ \mev \left( \frac{10^{-7}\ev}{m_\acc} \right)^2 \left(\frac{10^{15}\gev}{M_*}\right).
\ee
This would prevent the sterile state mixing with $\mu$ and $\tau$
neutrinos from being populated until after $\nu_{\mu,\tau}$ production
had ceased. The couplings to the electron, however, would still be essential to prevent the sterile states mixing strongly with $\nu_e$ from thermalizing. Such a model could arise in extra dimensional scenarios, for instance, where the acceleron was dominantly localized on the muon brane.

Ultimately, if we are to consider multiple sterile neutrinos instead of just one light one, we are forced to consider as well certain tensions. However, the tensions are not severe, and straightforward solutions exist.

\subsection{Initial Conditions and Acceleron Dark Matter}

We have assumed up to this point that the acceleron lies at the minimum of its potential, but it could have begun with energy in the form of coherent oscillations. There are a number issues we must consider. The first is that of the usual moduli problem, that the energy in acceleron oscillations not cause early matter domination.  This implies
\be
\frac{1}{2}m_\acc^2 \langle \acc_i^2\rangle \left(\frac{T_{rec}}{T_i}\right)^3 < \rho_{rec},
\label{acci}
\ee
where $T_i$, the temperature where \acc\ begins to oscillate, is defined by the condition $m_\acc = H(T_i)$, and $\acc_i$ is the initial acceleron vev, which we assume can be very large.  Eqn.~\ref{acci} then yields the condition

\be
\langle \acc_i^2\rangle \lesssim (10^{12} \gev)^2 \left(\frac{10^{-6} \ev}{m_\acc}\right)^{1/2},
\ee
so that for a wide range of $\acc_i$, the energy in acceleron oscillations does not exceed the total energy in dark matter.

A second concern is that prior to BBN, when the neutrino scattering
and annihilation processes occur at a rate $\gamma$ which exceeds the
oscillation rate of the acceleron about its minimum, $\gamma > m_\acc$, it is possible that for certain periods of time we will have $m_n < T$, allowing the sterile states to be thermalized. Prior to electroweak symmetry breaking (EWSB), the Yukawa coupling between the sterile neutrino, active neutrino and Higgs is too small to allow thermalization. Hence we are only worried about thermalization during oscillation after EWSB.  Should this occur, there can be large contributions to the acceleron effective potential, driving $\vev{\acc}$ towards smaller values, allowing complete thermalization to occur.

The amount of time that the oscillating field stays below $\lambda \acc < T$ for $\vev{\acc^2} \gg \vev{\acc}^2$ can be estimated as
\be
\frac{T}{\lambda \dot \acc} \approx \frac{T}{\lambda m_a \vev{\acc^2}^{1/2}}
\ee

The scattering rate is $\sim 10^{-22} (T/\mev)^5$. Hence, for an initial vev of $10^{12}\ \gev$ or less, we see that it is inevitable that there will be many scatters during the period when the sterile state can be thermalized.

We have already seen that the finite temperature expectation value for \acc\ falls like $T^2$, while the amplitude of oscillations $\vev{\acc^2}$ falls like $T^3$, so this requirement becomes more stringent at later times. Once $m_\acc > \tau^{-1}$ we can just use the average value of $\vev{\acc}$, since there is insufficient time to thermalize. Requiring that the amplitude of oscillation is smaller than the expectation value at $T_{x}$ when $\tau^{-1}=m_\acc$ gives us an initial amplitude

\be
\vev{A_i^2} < \frac{m_e^4 T_i^3 T_{x} \alpha^2_e}{36 m_\acc^4 \mpl^2}
 \simeq  10^{10}\beta_e^2  \left(\frac{10^{-6}\ \ev}{m_\acc}\right)^{23/10}\gev^2.
\ee

Satisfying this constraint ensures that there is no conflict from BBN. However, if we additionally want to ensure that the oscillations do not spoil the initial conditions for neutrino dark energy, we must concern ourselves with what occurs at somewhat later times. After BBN, electrons become non-relativistic and quickly lose any significant contribution to the effective potential. At this point, only the neutrino contributions to the acceleron potential are relevant.

If the acceleron can oscillate all the way to zero from this minimum, then it is possible that the neutrinos hop from one mass eigenstate to the other when they are degenerate (i.e., when $\acc=0$). This is particularly possible because the mixing terms go to zero more slowly than the difference of the diagonal eigenvalues, leading to a particularly non-adiabatic situation.

At small values for \acc, the contribution to the potential from the neutrinos reaches a maximum of $T^2 m_D^2/24$. If there is less energy in the oscillations at this point than this, the acceleron will never oscillate to zero. Thus, we have a limit on the initial amplitude
\be
\frac{m_D^2 T_e^2}{12}> m_\acc^2 \vev{\acc_i^2} \left(\frac{T_e}{T_i}\right)^3.
\ee
Taking $T_e=m_e$, we find 
\be
\vev{A_i^2}< 10^{15}
\left(\frac{10^{-6}\ev}{m_\acc}\right)^{1/2} \left(\frac{m_D}{10^{-2} \ev}\right)^{2} \gev^2.
\ee

\section{Non-Standard Matter Effects}

The lower limit on $\beta_e$ indicated in eqn.~\ref{couplinglimits}
gives rise to the possibility of experimentally verifiable signals in
neutrino oscillation experiments.  In \cite{Kaplan:2004dq}, it was shown that
these new matter effects, for typical earth densities could overwhelm
the traditional MSW effect in earth and give rise to significant
signals in neutrino oscillation experiments.  The new matter effects
give rise to an effective potential
\be
V(\acc) = \frac{\beta_e}{M_{pl}}\rho_e\acc + \frac{1}{2}m_\acc^2
  \acc^2,
\ee
where $\rho_e \approx 10^{-3}\mbox{g/cm}^3$ is the typical electron density in earth.

Then the sterile neutrino mass will change in earth as \be
\Delta m_n = \lambda \Delta \acc \sim 10^{-1} \ev \times \lambda\ \beta_e \left(\frac{10^{-6} \ev}{m_\acc}\right)^2.
\label{deltamn}
\ee
The requirement that mass eigenstates be populated to produce dark
energy at late times, however, gives a lower
limit on the product $\lambda \beta_e/m_\acc^2$ through eqn.~\ref{couplinglimits}.  Combining with eqn.~\ref{couplinglimits} (recognizing that we must evaluate the mass here at zero temperature), we
find a minimal expected size of matter effects which is \be
\Delta m_n \gtrsim 10^{-1} \ev.
\ee
If the acceleron also couples to baryons, the effect could be much stronger. Since forces couping to baryon number do not violate the equivalence principle as strongly as those coupling to electron number, the effects from $\alpha_B$ could be roughly fifty times larger. We emphasize that the above value is merely a lower limit on this effect.

Thus one expects matter effects to have significant impact in
experiments which may constrain the sterile neutrino mass and mixings, for example LSND \cite{Athanassopoulos:1998pv}, KARMEN \cite{Armbruster:2002mp} and
MiniBooNE \cite{McGregor:2003ds}.  It has been show that such sensitive matter effects can help to reconcile the LSND
result with the Bugey \cite{Declais:1994su}, CDHS \cite{Wotschack:1984ny} and KARMEN experiments, where the
combination of limits from the experiments disfavors 3+1 and 2+2 fits
to the neutrino oscillation data \cite{Zurek:2004vd}.  In addition, such
large matter effects may modify MiniBooNE's sensitivity when compared with LSND.

We expect smaller, though perhaps not insignificant, effects on the mostly
active mass eigenstates: \be
\Delta m_\nu = \Delta m_n \frac{m_D^2}{m_n(m_n+\Delta m_n)},
\ee where $m_n$ is the vacuum sterile neutrino mass.

\subsection{Consequences for future neutrino experiments}
We have shown that requiring a set of initial conditions which satisfy BBN constraints and give the appropriate initial conditions for dark energy at late time give rise naturally to non-standard matter effects of a size which could be detectable at upcoming experiments.

One interesting possibility, noted elsewhere \cite{Kaplan:2004dq,Zurek:2004vd}, is that such a matter effect might serve to explain the discrepancy between LSND and the null short-baseline experiments. In particular, because the pathlength at the Bugey experiment is mostly air, while LSND is mostly earth, the allowed parameter space for the LSND signal is significantly widened. This would then allow a signal at MiniBooNE, even in a region of the parameter space which is already excluded in the standard 3+1 scenario. Such a signal might reasonably not agree with the LSND results, if the intervening material is sufficiently different.

Future short baseline experiments also hold the possibility to detect this signal. In short baseline experiments, the amplitude for oscillation goes as
\bea
1-P_{ee}=  \sin^2 2 \theta \sin^2(\Delta m^2 L/4 E) \simeq \frac{m_n^2 L^2 m_D^2}{4 E^2}\hskip 0.15in  \\ \nonumber = .016 \times \left( \frac{m_n}{1\ \ev}\right)^2 \left( \frac{m_D}{10^{-1}\ \ev} \right)^2 \left( \frac{L}{50\ {\rm m}} \right)^2 \left( \frac{100\ \mev}{E} \right)^2 \eea
where the second equality holds in the limit that $m_D \ll m_n$ and $L
< 4 E/m_n^2$. Hence we see that should $m_n$ go up in matter, the
effect would be to increase the oscillation probability into sterile
states. An experiment with the sensitivity of 
MiniBooNE may detect such an effect with $m_D \sim 0.2\ \ev$ and $\delta m_n \sim 1\ \ev$. Alternatively, a two-detector experiment to search for $\theta_{13}$, might perform two high statistics runs, with different amounts of material between the neutrino source and the near detector, and thus place limits on this effect.

\section{Summary}
The past decade has been extremely exciting for beyond the standard model physics, in that we have now found convincing evidence for neutrino mass and dark energy. A possible connection between the two requires the introduction of a singlet neutrino and a new scalar field, and thus forces us to consider their properties in the early universe.

Couplings of the scalar  to electrons
with gravitational strength are to be naturally expected. We have shown such interactions significantly alter the early
universe cosmology of the system.  In particular, the potential
created by this coupling generally forces the singlet neutrino to be sufficiently heavy that it is thermally inaccessible through BBN, when neutrino scattering and
annihilation processes which can populate the sterile state decouple.

A sufficiently large coupling would naturally lead to significant effects at terrestrial neutrino experiments, particularly with regard to the mass of the singlet neutrino. Such effects change the relative sensitivity of different experiments depending on the electron density of the medium of neutrino propagation.

The process by which the singlet state is thermally inaccessibly at early times also leaves the relic neutrinos in mass eigenstates at late times. This is important as it provides the proper initial conditions for recently considered ``hybrid'' models of neutrino dark energy. Such an initial condition is generic with these couplings, and should be a motivation to consider new matter effects a natural prediction of neutrino dark energy theories, possibly detectable at MiniBooNE or a double-CHOOZ style experiment.

\vskip 0.15in

\noindent {\bf Note added:} As this work was to be submitted, \cite{added} appeared, which considers the consequences of MiniBooNE for MaVaN scenarios.

\section*{Acknowledgements} The work of N. Weiner was supported by NSF CAREER grant PHY-0449818. The work of K.~Zurek was supported by DE-FG03-00ER41132. We thank A.~Nelson and E.~Adelberger for useful discussions.
\bibliography{early}
\bibliographystyle{apsrev}

\end{document}